\documentstyle[12pt]{article}
\renewcommand{\baselinestretch}{1.4}
\newcommand{\be}[1]{\begin{equation}\label{#1}}
\newcommand{\ee}{\end{equation}}
\newcommand{\ba}[1]{\begin{eqnarray}\label{#1}}
\newcommand{\ea}{\end{eqnarray}}
\newcommand{\rf}[1]{(\ref{#1})}

\newcommand{\text}[3]{\mbox{\hspace{#1em}#2\hspace{#3em}}}
\newcommand{\R}{ \mbox{\rm I$\!$R} }

\title{Multidimensional Topological Foam}
\author{
A.Zhuk\thanks{
Permanent address:Department of Physics, University of Odessa,
     2 Petra Velikogo Str., 270100 Odessa, Ukraine
} \thanks{
This work was supported in part by DFG grant 436 RUS 113/7/0
} \thanks{
E-mail:$<$zhuk@physik.fu-berlin.de$>$ and $<$zhuk@paco.odessa.ua$>$
}\\
Fachbereich Physik, Freie Universit\"{a}t Berlin,\\
Arnimallee 14, D-14195 Berlin, Germany\\
and\\
Institut  f\"{u}r Mathematik, Universit\"{a}t Potsdam,\\
Am Neuen Palais 10, D-14451 Potsdam, Germany
}

\date{}
\begin{document}
\maketitle%
\renewcommand{\baselinestretch}{1.4}
\abstract{
Multidimensional cosmological model with the topology
$M=\R\times M_1\times\ldots\times M_n$ where
$ M_i \ (i=1,\ldots ,n)$ undergo a
chain splitting into arbitrary number of compact spaces is
considered. It is shown that equations of motion can be
solved exactly because they depend only on the effective
curvatures and dimensions and "forget" about inner
topological structure. It is proved that effective
cosmological action for the model with $n=1$ in the case of
infinite splitting of the internal space coincides with the
tree-level effective action for a bosonic string.
}
%
%
\newpage

\section{Introduction}
\setcounter{equation}{0}

Our observable universe at the present time at large scales
is well described by the Friedmann model with 4-dimensional
Friedmann-Robertson-Walker (FRW) metric. However, it is possible
that spacetime at short (Planck) distances might have a dimensionality
more than four and possess complex topology. Quantum fluctuations in
metric at Planck distances might result in branching off or joining
onto baby universes. Both of these phenomena ( multidimensionality and
quantum fluctuations) provide a foamlike structure of the universe at
Planck length \cite{1,2}. In the present paper we shall consider a foamy
structure of the universe due to former of these phenomena. Namely,
we shall consider universe which has complicated topological structure
due to extra dimensions and this type of structure we call topological
foam.

At Planck length all dimensions (external and internal) have equal rights.
It might have place, for example, in the early universe at Planck times.
To describe dynamics of this universe it is natural to generalize the FRW
model to a multidimensional cosmological model (MCM) with the
topology \cite{3}
\be{1.1}
M=\R\times M_1\times\ldots\times M_n  ,
\ee
where $ M_i \ (i=1,\ldots ,n)$ denote $d_i$-dimensional
spaces of constant curvature. One of this spaces, say $M_1$,
describes our 3-dimensional external space but all others are internal
spaces. To make the internal dimensions unobservable at the present time the
internal spaces have to be reduced to scales near the
Planck length. This can be achieved by contraction of the
internal spaces during cosmological evolution or by supposing
that they are static (or nearly static) and of the Planck scales from the very
beginning. Anyway, the internal spaces have to be compact
spaces what can be achieved by
appropriate periodicity conditions for the coordinates
\cite{4,4a,5,6}. As a result, internal spaces may have a
nontrivial global topology , being compact (i.e. closed
and bounded) for any sign of the spatial curvature.
The model \rf{1.1} can be generalized
to the case of all spaces being Einstein spaces. Dynamics
of the factor spaces $M_i$ is described by scale factors
$a_i(\tau),\ (i=1.\ldots ,n)$. Now, we suppose that factor
spaces $M_i$ have structure of russian "matreshka":
each $M_i$ is split into a product of Einstein spaces
$M_i\to \prod_{k=1}^{n_i}M_i^k$. In one's turn each $M_i^k$
can be split also and so on. We demand only that factor
spaces corresponding to the most elementary splitting should
be Einstein spaces. In our paper we show that under such
conditions equations of motion can be solved exactly in spite
of very complex topological structure of the manifold $M$.
It takes place because scalar curvatures of the constituent
factor spaces are included into these equations in certain
combinations only. They can compensate each other provided
they have opposite signs.

Of special interest are exact solutions because they can be
used for a detailed study of the evolution of our space,
of the compactification of the internal spaces and of the
behaviour of matter fields. It was shown up to now that MCM
can be integrated if they have at the most two non-Ricci-flat
spaces $M_i$ \cite{7}-\cite{13}. Now, we can generalize all
these solutions to the cases with much more complex
topological structure. Similar ideas of splitting were
proposed also in \cite{12,13} for particular models.

We can continue the splitting process up to infinity.
Infinite splitting of the internal spaces corresponds to infinite
number of the internal dimensions. Scale factors in this case
should belong to the Banach space \cite{13}. In our paper we
show that in the limit of infinite splitting effective action
for the model with one internal space coincides with
tree-level effective action for a bosonic string in the
presence of a background gravitational field and dilaton
field.

%
%
\section{Model and splitting procedure}
\setcounter{equation}{0}

We consider a cosmological model with the metric
\be{2.1}
g=-e^{2\gamma(\tau)}d\tau\otimes d\tau+
e^{2\beta^1(\tau)}\sum_{k=1}^{n_1} g^{(1)}_{(k)}+
\ldots +
e^{2\beta^n(\tau)}\sum_{k=1}^{n_n} g^{(n)}_{(k)},
\ee
which is defined on the manifold \rf{1.1} where
\be{2.2}
M_i=\prod_{k=1}^{n_i}M_i^k,\  i=1,\ldots ,n.
\ee
The manifold $M_i^k$ with the metric $g^{(i)}_{(k)}$ is an Einstein
space of dimension $d_i^k$, i.e
\be{2.3}
R_{mn}\left[ g^{(i)}_{(k)}\right] = \lambda^i_k g^{(i)}_{(k)mn},\
m,n=1,\ldots ,d_i^k,
\ee
and
\be{2.4}
R\left[ g^{(i)}_{(k)}\right] = \lambda^i_k d^k_i.
\ee
The scalar curvature corresponding to the metric \rf{2.1} is
%
$$
R[g] = \sum^n_{i=1}e^{-2\beta^i}R\left[ g^{(i)}\right]
+e^{-2\gamma}\sum^n_{i,j=1}(d_i\delta_{ij}+d_i d_j)
\dot{\beta}^i\dot{\beta}^j
$$
\be{2.5}
+e^{-2\gamma}\sum^n_{i=1}d_i(2\ddot{\beta}^i -2\dot{\gamma}\dot{\beta}^i),
\ee
where
\be{2.6}
R\left[ g^{(i)}\right] = \sum_{k=1}^{n_i}R\left[ g^{(i)}_{(k)}\right]
\ee
and
\be{2.7}
d_i = \sum_{k=1}^{n_i}d_i^k.
\ee
Here,
$g^{(i)}=\sum_{k=1}^{n_i} g^{(i)}_{(k)}$ is the metric on
$d_i$-dimensional manifold $M_i$.
The overdot denotes differentiation with respect to time $\tau$.
It follows from the equation \rf{2.5} that the scalar curvatures and
dimensions of the constituent factor spaces $M_i^k$ are included there in the
combination \rf{2.6} and \rf{2.7} respectively.
It is not difficult to show that $M_i$ are also Einstein spaces provided
$M_i^k$ are not Ricci flat \cite{12,13}.
We can achieve it by conformal transformation
\be{2.8}
g^{(i)}_{(k)}\longrightarrow \left(
\lambda_k^i /\lambda^i
\right)
g^{(i)}_{(k)},
\ee
where $\lambda^i\neq 0 $ are arbitrary constants. After this transformation
$g^{(i)}=\sum^{n_i}_{k=1}\left( \lambda^i_k/\lambda^i\right) g^{(i)}_{(k)}$.
Using the equations \rf{2.3} and \rf{2.4} we get the following equations:
\be{2.9}
R_{mn}\left[ g^{(i)}\right] = \sum_{k=1}^{n_i}R_{mn}
\left[ \left(
\lambda_k^i /\lambda^i
\right) g^{(i)}_{(k)}\right]= \sum_{k=1}^{n_i}\lambda^i_k
g^{(i)}_{(k)mn}=
\lambda^i g^{(i)}_{mn},
\ee
\be{2.10}
R\left[ g^{(i)}\right] = \sum_{k=1}^{n_i}
\left(
\lambda^i /\lambda_k^i
\right)
R\left[ g^{(i)}_{(k)}
\right]
=\lambda^i d_i,
\ee
which prove our statement. It is clear that the metric
$g^{(i)}$ is indefinite now if signs of
$ \lambda^i, \lambda^i_1,\ldots , \lambda^i_{n_i}$
do not coincide  at least for two of them.

The action is adopted in the following form
\be{2.11}
S= S_g +S_m +S_{GH}=
\frac{1}{2\kappa^2}\int_M d^ x\sqrt{|g|}(R[g]-2\Lambda)+ S_m +S_{GH},
\ee
where $S_{GH}$ is the standard Gibbson-Hawking boundary term \cite{14},
$\kappa^2$ is $D=1+\sum_{i=1}^n d_i$-dimensional gravitational constant,
$\Lambda$ is a cosmological constant, and $S_m$ is action for matter. For the
metric \rf{2.1} the action \rf{2.11} reads
$$
S_g +S_{GH}=
\mu\int d\tau \left\{
e^{-\gamma+\gamma_0}G_{ij}\dot{\beta}^i \dot{\beta}^j -
\right.
$$
\be{2.12}
\left.
\frac12 e^{\gamma+\gamma_0}\left(
-\sum^n_{i=1} R\left[g^{(i)}\right] e^{-2\beta^i} +2\Lambda
\right)
\right\},
\ee
where $\gamma_0=\sum_1^n d_i \beta^i$. The effective curvatures
$R\left[g^{(i)}\right]$ and dimensions $d_i$ are defined by the equations
\rf{2.6} and \rf{2.7} respectively.
$\mu=\frac{1}{\kappa^2} \prod^n_{i=1}\mu_i$  where
$\mu_i=\prod^{n_i}_{k=1}\mu_i^k$ and
$\mu_i^k=\int_{M_i^k}d^{d_i^k}y\sqrt{|g^{(i)}_{(k)}|}$ is
the volume of $M_i^k$ (as was mentioned in the Introduction,
all $M^k_i$ are compact).  $G_{ij}=d_i\delta_{ij}-d_i d_j$
are the components of the minisuperspace metric.

In the case of homogeneous minimally coupled scalar field we
get
\be{2.13}
S_m=
\kappa^2\mu\int d\tau \left\{
\frac12 e^{-\gamma+\gamma_0}\dot{\varphi}^2 -
e^{\gamma+\gamma_0} U(\varphi)
\right\}.
\ee
We can consider also matter in the form of $m$-component
perfect fluid with the energy-momentum tensor
\be{2.14}
T_N^M=\sum^m_{a=1}T^{(a)M}_{\quad N},
\ee
\be{2.15}
T^{(a)M}_{\quad N}=diag\ \left(
-\rho^{(a)},\underbrace{P_1^{(a)},\ldots ,P_1^{(a)}}_{d_1},\ldots ,
\underbrace{P_n^{(a)},\ldots ,P_n^{(a)}}_{d_n}
\right) .
\ee
The conservation equations are imposed on each component
separately:
\be{2.16}
T^{(a)M}_{\quad N;M} =0\,\quad a=1,\ldots ,m
\ee
and for the tensor \rf{2.15} and the metric \rf{2.1} they
have the form
\be{2.17}
\dot{\rho}^{(a)}+\sum^n_{i=1} d_i
\dot{\beta}^{i}\left(
\rho^{(a)}+P_i^{(a)}
\right).
\ee

We suppose that the pressure and the energy density are
connected by the equations of state
\be{2.18}
P_i^{(a)}=\left(
\alpha_i^{(a)}-1
\right)
\rho^{(a)}
\ee
where $\alpha_i^{(a)}=const$. Then, the conservation equations \rf{2.17}
have the simple integrals
\be{2.19}
\rho^{(a)}(\tau)= A^{(a)}\prod_{i=1}^n a_i^{-d_i\alpha_i^{(a)}}=
A^{(a)}\prod_{i=1}^n V_i^{-\alpha_i^{(a)}},
\ee
where $A^{(a)}$ are constants of integration, $a_i=\exp \beta^i$
are scale factors and $V_i$ are proportional to the volume of
$M_i:\ V_i= a_i^{d_i}$. It is not difficult to verify that the Einstein
equations with the energy momentum \rf{2.14}-\rf{2.19} are equivalent to the
Euler-Lagrange equations for the action \rf{2.11} and \rf{2.12} where
\be{2.20}
S_m= -\kappa^2 \mu\int d\tau e^{\gamma+\gamma_0}\sum_{a=1}^m \rho^{(a)}.
\ee

It can be easily seen from the equations \rf{2.12}, \rf{2.13} and \rf{2.20}
that scalar curvatures $R[g^{(i)}_{(k)}]$ and dimensions $d_i^k$
of the constituent factor-spaces $M_i^k$ are included into action in
combinations \rf{2.6} and \rf{2.7} only. The equations of motion
(Euler-Lagrange equations) "forget" about topological structure. Thus,
all possible combinations of the spaces $M_i^k$ which give the same effective
curvatures $R[g^{(i)}]$ and the same effective dimensions $d_i$ will have the
same dynamics. We can use this fact to generalize the solutions obtained in
\cite{7}-\cite{13}. Let us consider, for example, the case
\be{2.21}
\sum^{n_i}_{k=1}R\left[
g^{(i)}_{(k)}
\right]=0,\ i=1,\ldots, n.
\ee
Then, for MCM in the presence of minimally coupled scalar field and
$\Lambda=0$ we obtain two classes of the solutions \cite{11}:\\
\noindent i) generalized Kasner solutions
\be{2.22}
a_i=a_{(0)i}t^{\alpha^i},\ i=1,\ldots ,n,
\ee
\be{2.23}
\varphi=\ln t^{\alpha^{n+1}}+ const,
\ee
where
\be{2.24}
\sum^n_{i=1}d_i\alpha^i=1,\quad \sum^n_{i=1}d_i(\alpha^i)^2=
1-(\alpha^{n+1})^2 .
\ee
Here, $t$ is synchronous time. The space volume of the universe is
proportional to time: $V\sim \prod^n_{i=1}a_i^{d_i}\sim t$.\\
\noindent ii) steady state solutions
\be{2.25}
a_i=a_{(0)i}\exp (b^i t),\ i=1,\ldots ,n,
\ee
\be{2.26}
\varphi= b^{n+1}t+ const,
\ee
where
\be{2.27}
\sum^n_{i=1}d_i b^i=0,\quad \sum^n_{i=1}d_i(b^i)^2 +(b^{n+1})^2=0.
\ee
For this class of solutions the space volume of the universe is constant:
$V=const$.

%
%
\section{Multidimensional cosmology and string theory}
\setcounter{equation}{0}

Let us slightly generalize the metric \rf{2.1} to the inhomogeneous case
supposing that the metric has the form
\be{3.1}
g=g^{(0)}+ e^{2\beta^1 (x)}\sum_{k=1}^{n_1}g_{(k)}^{(1)}+
\ldots +e^{2\beta^n (x)}\sum_{k=1}^{n_n}g_{(k)}^{(n)},
\ee
where the metric $g^{(0)}$ is defined on the $D_0$-dimensional manifold $M_0$
and $x$ are some coordinates of $M_0$. Full manifold $M$ is:
$M=M_0 \times \prod^n_{i=1}M_i$ where $M_i$ is defined by the equation
\rf{2.2} and all $M_i^k$ are again Einstein spaces (see eqs. \rf{2.3} and
\rf{2.4}). The scalar curvature is
$$
R[g] = R\left[
g^{(0)}\right]+
\sum^n_{i=1}e^{-2\beta^i}R\left[ g^{(i)}\right]
$$
\be{3.2}
-\sum^n_{i,j=1}(d_i\delta_{ij}+d_i d_j)
g^{(0)\mu\nu}\partial_\mu \beta^i\partial_\nu \beta^j-
2\sum^n_{i=1} d_i \Delta\left[g^{(0)}\right] \beta^i,
\ee
where $\Delta\left[g^{(0)}\right] $ is the Laplace-Beltrami operator
corresponding to $g^{(0)}$ and the effective curvatures
$R\left[ g^{(i)}\right]$
and dimensions $d_i$ are defined again by the equations \rf{2.6} and \rf{2.7}
respectively.

After dimensional reduction the effective cosmological action
is \cite{15}
$$
S=S_g +S_{GH}=\frac{1}{2\kappa_0^2}
\int_{M_0} d^{D_0}x \sqrt{|g^{(0)}|}\prod^n_{i=1}e^{d_i\beta^i}
$$
\be{3.3}
\cdot
\left\{
R\left[g^{(0)}\right]-\sum^n_{i,j=1}G_{ij}g^{(0)\mu\nu}
\partial_\mu \beta^i\partial_\nu \beta^j+
\sum^n_{i=1}R\left[g^{(i)}\right]e^{-2\beta^i}-2\Lambda
\right\}.
\ee
Here, $\kappa_0^2=\kappa^2/\prod^n_{i=1} \mu_i$ is
$D_0$-dimensional gravitational constant and
$G_{ij}=d_i\delta_{ij}-d_i d_j$ are the components of the
midisuperspace metric.

Let us consider in more detail the case of one internal
space: $n=1$. Then the action \rf{3.3} reads
$$
S=\frac{1}{2\kappa_0^2}
\int_{M_0} d^{D_0}x \sqrt{|g^{(0)}|}
e^{-2\Phi}
$$
\be{3.4}
\cdot
\left(
R\left[g^{(0)}\right]-
4\omega g^{(0)\mu\nu}\partial_\mu \Phi\partial_\nu \Phi+
R\left[g^{(1)}\right]e^{\frac{4}{d_1}\Phi}-2\Lambda
\right),
\ee
where the dilaton $\Phi$ is defined via the scale factor
$a_1(x)=\exp \beta^1(x)$ as follows $e^{-2\Phi}:=a_1^{d_1}$
and $\omega=\frac{1}{d_1}-1 < 0$
is the Brans-Dicke parameter. This action is written in the
Brans-Dicke frame. After conformal transformation
\be{3.5}
\hat{g}^{(0)}_{\mu\nu}=e^{-4\Phi/(D_0 -2)}g^{(0)}_{\mu\nu},
\ee
\be{3.6}
\varphi=\pm 2\left[
\omega+\frac{D_0-1}{D_0-2}
\right]^{1/2}\Phi
\ee
the action \rf{3.4} can be written in the Einstein frame as follows
$$
S=\frac{1}{2\kappa_0^2}
\int_{M_0} d^{D_0}x \sqrt{|g^{(0)}|}
$$
\be{3.7}
\cdot
\left\{
\hat{R}\left[\hat{g}^{(0)}\right]-
\hat{g}^{(0)\mu\nu}\partial_\mu \varphi\partial_\nu \varphi+
R\left[g^{(1)}\right] \exp(-2\lambda_{(1)c}\varphi)
-2\Lambda \exp(-2\lambda_{(2)c}\varphi)
\right\},
\ee
where the cosmological dilatonic coupling constants are
\be{3.8}
\lambda^2_{(1)c}=\frac{D-2}{d_1(D_0 -2)},
\quad \lambda^2_{(2)c}=\frac{d_1}{(D-2)(D_0-2)},
\ee
and $D=D_0+d_1$ is the dimension of the manifold $M$.

By analogy with the splitting $M^i\longrightarrow
\prod^{n_i}_{k=1}M_i^k$ we can split spaces $M_i^k$ into new
factor spaces:$M_i^k\longrightarrow\prod^{l_k}_{l=1}M_i^{k,l}$
where $M_i^{k,l}$ are arbitrary Einstein spaces. In one's
turn spaces $M_i^{k,l}$ can be split also and so on (chain
splitting). We demand only that spaces corresponding to the
most elementary splitting are Einstein spaces. We can
continue the splitting process up to infinity and under this
limit the actions \rf{2.12}, \rf{2.13}, \rf{2.20}, \rf{3.3},
\rf{3.4}, and \rf{3.7} are well defined provided that
\be{3.9}
\sum^n_{i=1}\left|R\left[g^{(i)}\right]\right| <+\infty,
\quad \sum^n_{i=1}d_i|\beta^i| <+\infty.
\ee
In this case $\beta=(\beta^i)$ belongs to the Banach space
with $l_1$-norm \cite{7,13}.

In particular case $n=1$, infinite splitting of the internal
space corresponds to the limit $d_1\to\infty$ and the action
\rf{3.7} reads
\be{3.10}
S=\frac{1}{2\kappa_0^2}
\int_{M_0} d^{D_0}x \sqrt{|\hat{g}^{(0)}|}
\left\{
\hat{R}\left[\hat{g}^{(0)}\right]-
\hat{g}^{(0)\mu\nu}\partial_\mu \varphi\partial_\nu \varphi
+C\exp (-2\lambda_s\varphi)
\right\},
\ee
where
\be{3.11}
\lambda^2_s:=\frac{1}{(D_0 -2)},
\quad C:=R\left[g^{(1)}\right]-2\Lambda.
\ee
This action coincides with tree-level effective action for a
bosonic string in the presence of a background metric
$g^{(0)}_{\mu \nu}$ and dilaton $\Phi$. Here, $\lambda_s$ is
the string dilatonic coupling constant and $C$ is the central
charge deficit \cite{16,17}.

In conclusion I would like to note that infinite chain splitting may
result in topological fractal structure of the multidimensional
universe. We  investigate this interesting problem in our
forthcoming paper \cite{18}.

\bigskip


{\bf Acknowledgements}

The work was supported in part by DFG grant 436 RUS 113/7/0.
I wish to thank Prof.Kleinert and Free University of Berlin
as well as the Projektgruppe Kosmologie at the Potsdam
University for their hospitality during preparation of this
paper. I am grateful to V.D.Ivashchuk, U.G\"{u}nther and
M.Rainer for useful discussions.

%
%

%
%
\end{document}